# FlexiGen: Stochastic Dataset Generator for Electric Vehicle Charging Energy Flexibility

Bernardo Cabral, Tiago Fonseca, Clarisse Sousa, Luis Lino Ferreira
INESC TEC/Polytechnic of Porto - School of Engineering
Porto, Portugal
{bemac, calof, cassa, llf}@isep.ipp.pt

*Abstract*— **Electric vehicles (EVs) and renewable energy sources (RES) are vital components of sustainable energy systems, yet their uncoordinated integration can pose substantial challenges to grid stability, such as unmanaged peak loads and energy balance issues. Vehicle-to-Grid (V2G), offer a promising solution to address these challenges by enabling bidirectional energy flow between EVs and the grid. As such, EVs can be used in advances Demand Response (DR) strategies to optimize energy use and mitigate the intermittency of renewable generation. To reach such advantages, optimization algorithms need data on EV energy flexibility, such as charging patterns and usage preferences. However, data collection remains constrained by challenges such as high costs, user engagement, data privacy concerns, and limited access to open-source datasets on EV energy flexibility. This paper presents FlexiGen an open-source stochastic dataset generator tool designed to overcome the data limitations in EV flexibility for V2G and V1G DR applications. FlexiGen generates synthetic datasets encompassing realist EV usage patterns, behaviours and flexibility scenarios for household and office routines. To generate these datasets, FlexiGen uses a series of configurable probabilistic variables, such as stochastic user routines, traffic conditions, charger types, car average electricity consumption and State of Charge (SoC). The generated datasets include an hourly routine with the EV State of connection, Destination Charger, Estimated Departure Time, Required SOC at Departure, Estimated Arrival Time, and Estimated SOC at Arrival. Accompanying this publication an example dataset is generated for 3 households with 1 EV each, and 1 office building with 3 EVs. The generated dataset is analyzed and discussed on the paper and published alongside the open-source code for FlexiGen tool.**

## I. INTRODUCTION

Despite the large, and still growing, adoption rates of Electric Vehicles (EVs) and Renewable Energy Sources (RES) due to a series of general perception, environmental benefits, and economic incentives, these technologies, individually, cannot fully guarantee their sustainability potential [1], [2]. On the contrary, they can even introduce a series of infrastructural, control, technological, and institutional challenges with negative impacts on the grid [3], [4], [5], [6], [7], [8]. An important aspect of it is related with the unmanaged peak loads resulting from uncoordinated EV charging with DERs production during peak-loads.

### A. Vehicle to Grid

Vehicle-to-grid (V2G), and simpler technologies such as Grid-To-Vehicle (V1G) have the potential to be used to offset this problem, and, as a result, have gained significant attention in recent years. Other forms of energy storage exist, such as traditional pumped hydro storage, and stationary power banks, however they are deemed as not sufficient to address the issue of large-scale intermittency and peak-load mismatch in a fully renewable power grid.

Particularly, as RES such as wind and solar become more prevalent [9], the challenge of intermittency becomes more significant. V2G research projects have already demonstrated the potential and feasibility of V2G technology. [10], [10], [11] [12], [13], [14]. These projects pointed to digitization and data as essential to make V2G and other innovative Demand Response (DR) a reality. In fact, to foster all the benefits of V2G innovations, algorithms such as Machine Learning (ML) [15], [16], Reinforcement Learning (RL), meta-heuristics, and optimal control are used and researched to optimize EV usage patterns and energy flexibility [17].

With V2G technology in place, an algorithm could decide based on the flexibility of the EV owner. For example, if an EV owner which does not have PV panels is planning to arrive home at 3 pm and leave at 8 am the next day (<u>usage time flexibility</u>), and wants its battery to be filled with at least 80% battery when leaving (<u>energy flexibility</u>), the algorithm can charge the EV during the afternoon with energy from the solar panels of the neighbors to 100% State of Charge (SOC), and discharge it later, to utmost until 80% SOC to be ready to leave in the next morning, so that the energy can be used at his own house (energy which is renewable and came from a cheaper source during the afternoon), or sold to the neighbors within the Renewable Energy Community (REC) [18], [19]. This strategy can be economically beneficial as it reduces its total EV energy cost, and is, at the same time, environmentally friendly and effective to balance the energy grid.

Both real-time and historical data were deemed important by researchers for the V2G use cases. If an EV is delayed in traffic and will arrive at a charging station later than expected, real-time data can be fed to the V2G system so that it can adjust the charging schedule optimization to ensure that the REC energy domain stays balanced and optimized. On the other hand, historical data can be used, for example, to simulate a REC with EVs and with that perform informed investment decisions when setting the V2G or REC, or train and deploy an RL algorithm.

These examples emphasize the crucial need for extensive datasets on the energy and usage flexibility of EVs, specially on energy flexibility for the EV owner, and time of usage of the EV, to effectively implement and test V2G applications. It also notes that data is also particularly important for learning-based algorithms like RL, which have demonstrated the most promising results in the literature and show the greatest potential for real-world implementation.

## B. Gaps and Contributions

Despite its value, several challenges remain in the collection and availability of data, specifically data on the usage and energy flexibility patterns for charging EVs, essential for V2G/V1G strategies as highlighted before. Data value for downstream services and analysis only becomes visible once it has been collected and analysed, sometimes for large periods of time. Moreover, when investing in such data collection projects seek confidentiality and do not disclose the data for open-source initiatives as it may represent a competitive advantage for future products, it may be sold to interested market parties, or it may violate GDPR legislation [20][21]. Other significant challenges for collecting data include operational issues such as the limitations of automation and user engagement, for example, for stating their energy and time flexibility.

Recognizing the highlighted problematics, researchers have been keen on comprehensively reviewing datasets and tools related to energy systems datasets or their generation. For example, [22] reviews open-source energy data with a focus on its utilization for energy communities. However, the authors do not further explore EVs data neither flexibility data, which is fundamental for the implementation of a V2G/V1G solutions. On the other hand, [23] focuses only on the EVs datasets, which by their own cannot be leveraged to develop an integrated V2G prepared for a real-world implementation. Dataset generation methods relevant to the one applied by this paper are discussed and analysed in the State of the Art Section II. However, to the best of the authors' knowledge none of the identified dataset generators, neither available open-source datasets encompass configurable energy and usage flexibility of EVs, essential for V2G/V1G simulation and real-world implementation.

On this notice, this paper presents FlexiGen, an open source stochastic dataset GEnerator for EV FLEXIbility datasets designed to overcome the data limitations in EV flexibility for V2G and V1G DR applications. FlexiGen generates synthetic datasets encompassing realist EV usage patterns, behaviours and flexibility scenarios for both household and office charging types of routines.

In contrast to the identified State of the Art (Section XXX), FlexiGen offers an approach by not only considering EV arrival and departure times but also incorporating data from census reports, such as traffic patterns, working hours, distance from work, and specific parameters related to charging infrastructure (e.g., charger power and battery capacity). This integration of additional layers of realism and variability distinguishes our tool from previous models. Moreover, FlexiGen has the capability to leverage demographic and behavioral data (configurable by the user), enabling the simulation of routines that are more closely aligned with real-world usage patterns. This includes variations in charging habits based on day types (e.g., weekdays vs. weekends), traffic conditions, and working hours, which adds a layer of context-sensitive behavior that is often overlooked in existing models. This configurability can be used to tailor datasets and simulations to specific regions or user groups, thereby enhancing the applicability of the generated data.

The generated datasets include an hourly routine with the EV State of connection, Destination Charger, Estimated Departure Time, Required SOC at Departure, Estimated Arrival Time, and Estimated SOC at Arrival.

To test FlexiGen, an example dataset is generated for 3 households with 1 EV each, and 1 office building with 3 EVs considering Portugal usage patterns and flexibility statistics. The generated dataset is analyzed and discussed at Section XXX. Both the dataset and the open-source code for FlexiGen are made available alongside the publication.

The content in the article is significant not only for engineers who are keen on designing, simulation or operationalizing V2G/V1G initiatives, but also for electric utilities, DER aggregators, and policymakers researching the smart grids and the effects these technologies can have. The main contributions of this article can be resumed as the (i) development of FlexiGen as an open-source stochastic dataset generator for EV flexibility, (ii) the discussion of challenges and future directions of collecting and generating synthetic datasets for EV flexibility, (iii) generating and analyzing an open-source dataset for EV flexibility.

## C. Outline

The remainder of the paper is structured as follows: Section 2 deepens the context and covers the State of the Art on dataset generators. Section 3 presents the assumptions made for developing FlexiGen and Section 4 presents the tool itself. Finally, Section 5 provides a description of the generated dataset to test FlexiGen and Section 6 concludes the work identifying future work perspectives.

## II. STATE OF THE ART

The synthetic generation of EV data is becoming increasingly important due to the growing adoption of EVs and the need for efficient grid management. Lahariya et al. present a Synthetic Data Generator (SDG) [24] that models EV charging sessions based on real-world data. Their approach emphasizes temporal modeling, specifically managing vehicle arrival and departure times across charging station networks using techniques like exponential distributions and Gaussian mixture models (GMMs).

Earlier works contributed to this area using various statistical and machine learning methods. For example, Flammini et al.[25] employed beta mixture models to statistically represent EV session arrival times, but their model lacked a temporal aspect for synthetic data generation. Additionally, studies like [26] applied generative adversarial networks (GANs) to model continuous energy consumption, while [27] used k-nearest neighbors (k-NN) for predicting charging demands at specific stations, though without accounting for session durations. Related to this, Gaete-Morales et al. introduced emobpy [28] , a tool for generating synthetic time series data for electric vehicles (EVs). This model incorporates EV session durations by tracking when vehicles are driving or parked, and whether they are connected to charging stations. Unlike emobpy, which emphasizes overall EV mobility and electricity consumption, FlexiGen generates synthetic datasets that capture EV flexibility by simulating charging scenarios in household and office environments.

## III. ASSUMPTIONS (LLF)

The development of FlexiGen is based on a set of guiding assumptions designed to align the generated data with realistic EV user behaviors and energy flexibility requirements. At the core of these assumptions is the concept of the Flex Offer (FO), introduced in [29], which represents a prosumer's willingness to adjust energy consumption to aid in grid balancing. This concept is especially relevant in V1G/V2G, where an EV can dynamically shift its charging and discharging activities to support energy stability.

In FlexiGen, we adopt the most flexible FO type, elastic offers [30], allowing for flexibility in both the timing of energy use and the amount of energy consumed or returned. This means that the EV's charging and discharging schedules can adapt to the grid's needs while meeting the owner's energy requirements. Specifically, we model each EV as having both time flexibility (from when to when it can be charged or discharged) and energy flexibility (how much energy it needs to store or can offer back to the grid), enabling more realistic and adaptable scenarios for energy management.

Our assumptions extend beyond energy flexibility to the behavior patterns of typical EV users. FlexiGen generates datasets based on assuming a profile of a commuter who leaves home in the morning, driving an EV to work over a defined distance. Although this departure time is largely predictable, we introduce stochastic elements to reflect day-to-day variations in routine. This means that while the person's schedule is mostly consistent, factors like minor changes in morning activities, or out of the normal events, are factored into the model. It is also possible to introduce small stochastic changes to the distance travelled.

/* assume-se que estão ligados ao carregador o dia todo? */

/* há mais carros ligados ao mesmo carregador */

/* quantas FO são geradas: 1 para a casa e outra para o trabalho? */

/* falta explicar que suporta diferentes tipos de carros,

On their commute, the EV's battery level is impacted by various factors, including traffic conditions, driving speed, and weather. These elements are also modeled stochastically by user inserted statistics in FlexiGen to capture the variability in energy consumption across different journeys. For example, a higher-than-usual traffic intensity will increase energy consumption due to prolonged idling or slower speeds, while lighter traffic would decrease it. In this way, FlexiGen mirrors the dynamic energy consumption patterns that an EV might experience.

Additionally, FlexiGen takes into account differences in routines between weekdays and weekends. Weekdays are mostly characterized by predictable schedules, such as commuting to work, while weekends introduce more variability in terms of destinations, travel times, and charging behavior. This approach provides a realistic foundation for simulating EV energy flexibility, catering to the distinctive characteristics of various days and usage contexts.

These behavioral and energy flexibility assumptions are embedded in FlexiGen's design, allowing customization to fit diverse geographic and demographic contexts. Users of FlexiGen can adjust variables related to traffic patterns, work schedules, charger types, and even the behavioral tendencies of specific regional populations, thus tailoring the tool to meet specific simulation or research needs. By combining these elements, FlexiGen aims to generate datasets that not only reflect realistic EV usage but also offer insights into energy flexibility for V2G and V1G applications.

## IV. FLEXIGEN – STOCHASTIC EV DATASET GENERATION

FlexiGen generates datasets that simulate EV usage patterns and energy flexibility scenarios, providing vital information for optimizing charging infrastructure and energy distribution.

This chapter offers an in-depth exploration of the FlexiGen tool, focusing on its overall design and architecture, individual components, customizable generation parameters, its capacity to simulate the different and complex dynamics of EV integration in both residential and office contexts, and how it can fit with an established simulator, like CityLearn [31].

### A. FlexiGen Architecture

FlexiGen's architecture is designed to be modular and extendable, allowing users to configure and modify different components based on the needs of a specific application. The tool is divided into three main components: **Data Collection**, **Dataset Generation Modules**, and **Dataset Output**.

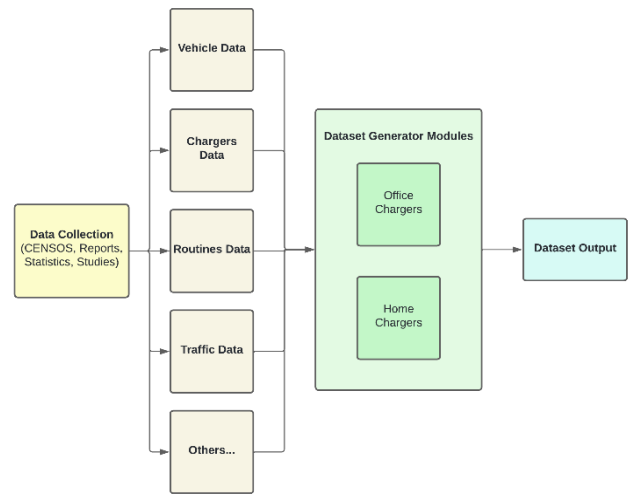

*Figure 1 – FlexiGen Architecture*

As an input, FlexiGen integrates several probabilistic variables to create diverse and realistic data drawing upon various census and other datasets studied on routines such as work schedules, traffic patterns, as well as the behaviour and information of electric cars. FlexiGen input is configurable, enabling the user to adjust input parameters and tailor dataset generation to match specific demographic and regional characteristics. Section III.B. goes into detail on considered variables and input simulation characteristics.

After inputting the variables, FlexiGen provides the user a choice between two distinct dataset generation modules: One for generating behaviours and routines for EVs used by household users; the second to generate flexibility datasets for EVs in an office or work-related buildings. Section III.C. specifies each generating module.

Finally, the datasets generated by FlexiGen contain temporal information, such as the times when an EV is connected to a charger, estimated departure and arrival times, required SOC at departure, and estimated SOC at arrival. Section III.D. goes into detail on output dataset description.

## B. Data Collection

The **Data Collection** component of FlexiGen is responsible for defining the context and configurability of the generated dataset. The input parameters are user customizable to suit different regions, vehicle types, charging infrastructure, and user profiles. These inputs define the operational and probabilistic conditions under which the synthetic datasets are generated.

- **Routine Variables:** Different user routines (weekdays vs. weekends) are configured to simulate real-life EV usage. The tool distinguishes between routines for office charging and home charging, allowing for realistic variations in user behavior based on work schedules and other factors.

- **Traffic Variables:** Traffic intensity during commuting hours is modeled to determine variations in travel times and energy consumption. This includes the probability of traffic congestion and its effect on SOC.

- **Charger and Vehicle Data:** Configurable information about charger types, charging rates, battery capacity, and energy consumption is provided to simulate different charging scenarios and infrastructure capabilities.

- **Probabilistic Parameters:** Several stochastic elements, such as probabilities of routine change, decision to charge, and weekend behavior, are used to introduce randomness into the generated dataset, improving its realism.

Table 1 summarizes the input data needed for FlexiGen.

*Table 1 – Home parameters*

| Variable Name | Description |
|---|---|
| ROTINA_CHANGE | Possibility of a routine change |
| MAX_BATTERY_CAPACITY | Max SoC Charge possibility |
| CHARGER_CHANGE | Probability of charging during travel |
| YEARS | Number of years for data generation |
| CHARGER_EX | Cars, chargers' id's, battery and charging capacity |
| DAY_WEEK | Leaving home to work routines / schedules |
| NIGHT_WEEK | Arriving home from work routines / schedules |
| WEEKENDS | Weekends routines /schedules / probabilities |
| DIST | Week distance from home or work places |
| DIST_WEEKEND | Weekend distance from hobbies/places/work |
| TRAFFIC_WEEK | Week traffic / probabilities / |
| TRAFFIC_WEEKEND | Weekend traffic / probabilities / minimum and maximum time |
| WORK_WEEKEND_CONSTANT | Probability of normally working on weekends |
| WORK_WEEKEND_RAND_1 | Probability of working on a random Saturday |
| WORK_WEEKEND_RAND_2 | Probability of working on a random Sunday |
| CHARGE_BAT | Minimum and maximum car charging levels |
| CARS | Cars and their attributes such as battery capacity and consumption |
| CHARGERS_ALL | All available chargers and their charging capacity |

## C. Dataset Generation Modules

The **Dataset Generation Modules** are the core of the FlexiGen tool. They use the input data to generate the datasets based on configurable stochastic models and programmed logic. The dataset generation process is divided into two main modes, that the user can choose from, focusing on different use cases: **Home Charging** and **Office Charging**.

### 1) Home

The **Home Charging Module** simulates residential EV charging behavior. This module generates a dataset for one EV at a time. The length of the dataset can be pre-defined by the user of FlexiGen. The produced datasets will vary based on various variables and probabilities, introduced in the Data Collection phase according to the following main points:

- **Routine Simulation:** The module simulates a daily work week routine where an EV leaves home in the morning, travels to a destination (work), and returns in the evening. Probabilities are assigned to the likelihood of what time one typically leaves home, and what time one typically leaves work, with variations to account for deviations from routine (such as work from home days or needing to go to a medical consult for example).

- **Weekend Behavior:** On weekends, the home charging behavior becomes more diverse. The tool simulates various activities, such as trips for leisure, shopping, or other hobbies, each associated with its own energy consumption and flexibility dynamics.

- **Charging Preferences:** The probability of charging the vehicle during the day at another location (not at home), such as a public charging station or at work, is modeled based on user preferences and infrastructure availability. This affects EV SoC and impacts the SoC flexibility an EV will have when connecting to the home charger.

- **Traffic Impact:** Traffic conditions are incorporated into the model, influencing travel times, energy consumption,

and arrival times, which directly impacts the SOC and the timing of charging events.

### 2) Office

The **Office Charging Module** simulates the charging behavior of EVs at an office or work-related charger locations. This mode is particularly useful for understanding the impact of shared charging infrastructure in office environments, especially when the specific organization has one or multiple charging points.

- **Commuting Simulation:** The tool models the commuting behavior of EVs from home to the office and back. It takes into account arrival and departure times, which are influenced by user routines, traffic patterns, and work schedules.

- **Weekend Shifts:** Although office charging activity is generally lower on weekends, the module accounts for exceptions, such as employees working weekend shifts or visiting the office for specific purposes.

- **Charging Scheduling:** The decision to charge at the office is probabilistic, influenced by factors like SOC upon arrival, the expected duration of stay, and the availability of chargers. This introduces complexity similar to real-world office charging scenarios, where not all vehicles can charge simultaneously.

- **Traffic Impact:** Similar to the previous module, traffic conditions are incorporated into the model, influencing travel times, energy consumption, and arrival times, which directly impacts the SoC and the timing of charging events.

### D. Output Datasets

The dataset generated by FlexiGen tool are structured into multiple Comma Separated Values (.csv) files, each corresponding to the an individual EV profile with energy flexibility. These files are organized by columns and rows. Each row represents a specific hour of a given day, and the columns contain different categories of data. The detailed description of each column is provided below:

- **Month**: This column records the month of the year, represented as an integer ranging from 1 (January) to 12 (December), to provide a chronological context for the generated data.

- **Hour**: The Hour column denotes the hour of the day, ranging from 1 to 24, thereby offering a time-based reference for each entry.

- **Day Type**: This column categorizes the day of the week numerically, from 1 (Monday) to 7 (Sunday). Additionally, the value 8 is used to indicate special days such as holidays, which may cause unique charging behaviors or different patterns in flexibility usage.

- **EV State**: This column describes the current state of the EV. It provides an indication of whether the vehicle is parked, connected to the "Charger" and ready to charge (denoted by 1) or in transit (denoted by 3). This classification is essential for determining the EV's energy flexibility at any given point in time. A special state, denoted by 2, can be used to signify that the EV is incoming to the "Charger", simulating a possible integration of a GPS software with the V2G system, which could provide such information.

- **Charger**: The Charger column specifies the identification of the charging station that the EV is connected or is incoming to. If no specific charging station is assigned, the value is represented as 'nan'. Otherwise, the charger ID is given in the format $EVC_{b\_n\_p}$ (where $EVC$ stands for Electric Vehicle Charger, $b$ stands for the Building where the charger is inserted in the simulation, $n$ stands for the charger Number within the building and $p$ stands as the number of the Plug of that charger). In EvLearn module or CityLearn this will facilitate the appropriate linkage between the EVC and the EVs during the simulation.

- **Estimated Departure Time**: This column provides the time flexibility of the EV owner with the estimated number of time steps until the vehicle is expected to depart from its parked and connected state. This information is relevant only when the EV is in the parked, connected and ready to charge state (State 1), offering insight into the duration available for energy exchange with the grid.

- **Required SOC at Departure**: The Required SOC at Departure column denotes the energy flexibility as a SOC percentage that the EV must achieve by its departure time, ranging from 0% to 100%. This is denoted by the EV owner and corresponds to its energy flexibility. This requirement is crucial for balancing energy flexibility and ensuring that user needs and comfort are met while participating in V2G services.

- **Estimated Arrival Time**: The Estimated Arrival Time column indicates the expected number of time steps until the vehicle arrives at the charging station. This information is available only when the EV is incoming (State 2) and ranges from 1 to 24, reflecting the expected arrival time within the following hours.

- **Estimated SOC at Arrival**: This column provides the projected SOC percentage of the EV upon arrival at the charging station, ranging from 0% to 100%. It is accessible only when the EV is incoming state (State 2) and helps in estimating the energy requirements for subsequent charging activities.

The structure of these files and the detailed data provided will enable comprehensive analysis and modeling of EV behavior, supporting both energy system planning and the development of V2G/V1G algorithms. The generated data aims to capture the intricacies of EV flexibility in terms of energy consumption, availability, and charging behavior, thus laying the foundation for advanced optimization and control strategies in energy systems. As discussed before, these files can be used as a standalone or used within CityLearn or EVLearn to simulate an REC DR system with V2G available.

### E. Integration within CityLearn

FlexiGen was developed to address a significant gap in the availability of datasets for energy flexibility modeling. This gap was identified during the development of a simulation tool, EvLearn, which is designed as an extension to simulate Electric Vehicle (EV) Vehicle-to-Grid (V2G) and Grid-to-Vehicle (V1G) interactions within an established simulation framework for Energy Management Systems (EMSs), namely the CityLearn platform. CityLearn, which already models various energy assets including buildings and renewable energy systems, lacked a comprehensive representation of EV flexibility—an essential component for accurately simulating advanced energy management scenarios involving EVs.

During the development of EvLearn, it became evident that the absence of detailed flexibility data hindered the effective modeling of V2G and V1G strategies. This realization led to the creation of FlexiGen. The primary objective of FlexiGen is to generate synthetic datasets that capture EV flexibility in a realistic manner, allowing researchers to use these datasets either as a standalone resource or as part of integrated energy simulations.

A key strength of FlexiGen lies in that it us already integrated within the CityLearn framework [31] and the EvLearn extension [32]. The generated dataset files are formatted to be compatible with CityLearn and EvLearn, facilitating direct integration without the need for extensive data preprocessing or reformatting. This compatibility ensures that researchers can leverage the generated data efficiently within the broader simulation ecosystem of CityLearn. Furthermore, the datasets produced by FlexiGen can be utilized not only for standalone V2G and V1G studies but also in simulations that involve other energy assets such as residential buildings, renewable energy communities (RECs), and distributed energy resources (DERs).

This level of integration enables the evaluation of demand response (DR) strategies and energy flexibility solutions across a wide range of scenarios. Researchers can investigate how EV flexibility contributes to optimizing energy use, balancing the grid, and reducing peak demand in synergy with other assets. The comprehensive integration of FlexiGen with CityLearn thus facilitates the extraction of meaningful insights regarding demand response optimizations and enhances the overall capacity to model and simulate sustainable energy solutions.

Datasets generated by FlexiGen are already available at the official open-source integration of CityLearn and have been used for the generation of results and research within DR algorithms, such as in the work of [33].

## V. EXAMPLES

To demonstrate the capabilities of FlexiGen, a sample dataset was generated for a scenario involving 3 households with 3 EVs and 1 office building with 3 EVs, usage patterns and flexibility statistics. This dataset provides insights into the daily routines of EV users, including charging events, SOC changes, and the allocation of chargers in an office environment.

### 1) Experiment Variables

Regarding the configured input variables, for the generated data related to routines, the probabilities were distributed as follows. The array related to arrivals at the work charger or departures from the home charger on regular weekdays is shown in Tables 1 and .2, corresponding to the input array from the previous chapter, "DAY_WEEK".

*Table 1 2 – Day week Office*

|  | Probability | Hour Min | Hour Max |
|---|---|---|---|
| Routine 1 | 40% | 7am | 8am |
| Routine 2 | 50% | 8am | 9am |
| Routine 3 | 5% | 5am | 6am |
| Routine 4 | 5% | 9am | 14am |

*Table 2 3 – Day week Home*

|  | Probability | Hour Min | Hour Max |
|---|---|---|---|
| Routine 1 | 40% | 6am | 7am |
| Routine 2 | 50% | 7am | 9am |
| Routine 3 | 5% | 9am | 10am |
| Routine 4 | 5% | 10am | 12am |

As for the routines of arriving at the home charger or leaving the work charger, these probabilities, related to the variable "NIGHT_WEEK" (Tables 3 and 4).

*Table 3 4 – Night week Office*

|  | Probability | Hour Min | Hour Max |
|---|---|---|---|
| Routine 1 | 40% | 5pm | 6pm |
| Routine 2 | 45% | 6pm | 7pm |
| Routine 3 | 5% | 3pm | 4pm |
| Routine 4 | 10% | 8pm | 11pm |

*Table 4 5 – Night week Home*

|  | Probability | Hour Min | Hour Max |
|---|---|---|---|
| Routine 1 | 87% | 5pm | 7pm |
| Routine 2 | 3% | 3pm | 4pm |
| Routine 3 | 10% | 8pm | 11pm |

For the distance associated with weekday routines, such as the commute distance, linked to the variable "DIST_HOME", it was configured as shown in Table .5 for demonstration purposes.

*Table 5 6 - Work Distance*

|  | Probability | Min Distance | Max Distance |
|---|---|---|---|
| Distance 1 | 25% | 6km | 10km |
| Distance 2 | 60% | 10km | 50km |
| Distance 3 | 15% | 60km | 90km |

Finally, there are other minor influencing factors, such as variables related to routine changes ("ROUTINE_CHANGE"), whether to charge during a trip ("CHARGER_CHANGE"), and the possibility of encountering traffic during the trip, both on weekdays ("TRAFFIC_WEEK") and weekends ("TRAFFIC_WEEKEND"), as detailed in Tables .6 and .7. In the Two tables, it's possible to observe the different probabilities of occurring different types of traffic, in each of them it's specified the probability, and the impact on the extra trip time.

*Table 6 7 – Traffic Week*

|  | Probability | Min Increase | Max Increase |
|---|---|---|---|
| Traffic 1 | 3% | 0% | 9% |
| Traffic 2 | 20% | 10% | 30% |
| Traffic 3 | 50% | 30% | 70% |
| Traffic 4 | 17% | 70% | 200% |

*Table 7 8 – Traffic Weekend*

| | Probability | Min Increase | Max Increase |
|---|---|---|---|
| Traffic 1 | 10% | 0% | 9% |
| Traffic 2 | 40% | 10% | 30% |
| Traffic 3 | 40% | 30% | 70% |
| Traffic 4 | 10% | 70% | 200% |

*2) Data Analysis*

After generating the datasets for the different cars and routines, as shown in Figure 2, the charging patterns at home chargers are inversely proportional to those at work chargers. This effect occurs because cars will charge at home when they are not at work, and at work, they will charge there if possible. For both home and work routines, slight differences can be observed, which could be attributed to variations in the probabilities for arrival and departure times, or even to different levels and durations of traffic encountered during the trips.

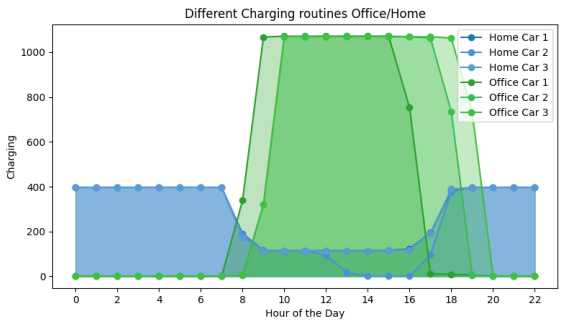

*Figure 2 Charging differences based on routines*

Regarding the routines, it is important to highlight that each time interval in the dataset is measured hourly. Thus, it is also possible to observe in the graph. ? and ?, the different travel times from home to work across four datasets. In most cases, trips take between 10 to 60 minutes. However, since the time intervals are hourly, we cannot distinguish finer details in these cases. Nonetheless, we can see that on certain days, due to factors such as heavy traffic or spontaneous changes in routines, there are notable shifts that result in longer travel times. It is worth noting that the script itself already includes some variations in travel times beyond just traffic and routine changes.

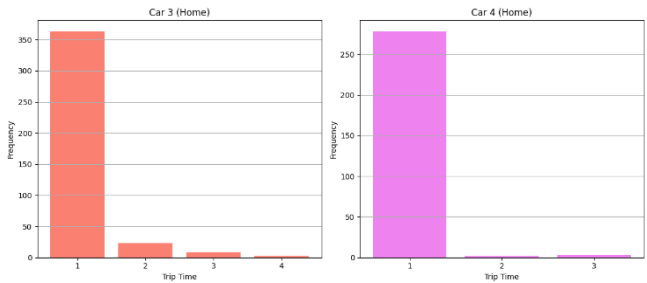

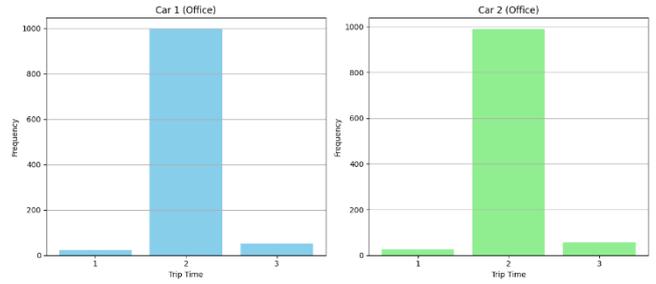

Regarding the time EVs remain connected to chargers, even if they are not actively charging, it is possible to distinguish two types of graphs: one for home chargers (Fig. ?) and another for workplace chargers (Fig. ?). For workplace chargers, it is notable that most charging sessions align with working hours, typically assuming arrival in the morning. This makes sense, as the majority of workers tend to work between 8 to 12 hours per day.

On the other hand, home chargers show longer connection times, as cars often remain plugged in for extended periods. Even though weekend activities or hobbies may cause some variation, in most cases, cars stay connected for longer durations, sometimes exceeding 60 hours. This suggests that, in some cases, vehicles are left plugged in from Friday until Monday. In other instances, it is common for cars to remain connected overnight after being plugged in after work.

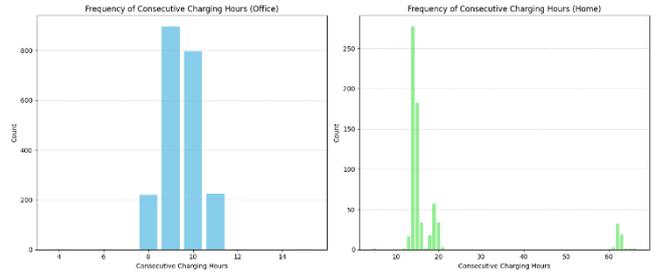

## VI. CONCLUSIONS

This paper introduced FlexiGen, an open-source tool for generating synthetic datasets that capture EV charging energy flexibility. Designed to address critical gaps in data availability, FlexiGen enables the simulation of EV usage patterns in both household and office settings, providing a a resource for the development and testing of DR strategies, such as V1G and V2G. FlexiGen's modular architecture and configuration allow users to input real-world probabilistic parameters—such as traffic conditions, departure times, and charging preferences based on their reality context—producing datasets that closely mimic realistic EV behaviors and charging scenarios. The generated datasets include time-stamped information on EV state of connection, estimated departure and arrival times, and SOC levels, supporting a comprehensive analysis of energy flexibility and offering a robust foundation for demand response (DR) optimization.

The generated datasets have been validated to align with typical EV user routines and provide insights into charging behaviors under different conditions. These datasets, in conjunction with FlexiGen's integration compatibility with CityLearn and EvLearn, provide a unique advantage by allowing researchers to simulate and optimize EVs within a broader scope of RECs. This is particularly valuable for electric utilities, companies, researchers, aggregators, and policymakers seeking to understand and leverage the impact of EVs on the grid.

Despite these contributions, FlexiGen has limitations that can be addressed in future work. The current version relies on static probabilistic parameters, which, while useful, may lack the dynamism required for more complex simulations involving real-time interactions. Future versions could benefit from integration with specialized EV simulation tools like SUMO (Simulation of Urban Mobility) [34], which would enhance FlexiGen's ability to model detailed traffic patterns and EV interactions with greater spatial and temporal resolution. Additionally, incorporating machine learning models to dynamically adjust probabilities based on historical data could further improve the accuracy and realism of the generated datasets.

Future work could also explore enhancing FlexiGen's configurability by introducing additional behavioral parameters, such as seasonal changes in traffic patterns or varying energy flexibility levels across different user demographics. With future enhancements, FlexiGen can further bridge the gap between synthetic data generation and real-world energy management applications, helping to unlock the full potential of V1G/V2G technologies in sustainable energy systems,, and ultimately contribute to a more sustainable future of transportation and energy generation, management and consumption.

## ACKNOWLEDGMENT

This paper is supported by the OPEVA project that has received funding within the Chips Joint Undertaking (Chips JU) from the European Union's Horizon Europe Programme and the National Authorities (France, Czechia, Italy, Portugal, Turkey, Switzerland), under grant agreement 101097267. The paper is also supported by Arrowhead PVN, proposal number 101097257. Views and opinions expressed are however those of the author(s) only and do not necessarily reflect those of the European Union or Chips JU. Neither the European Union nor the granting authority can be held responsible for them. The work in this paper is also partially financed by National Funds through the Portuguese funding agency, FCT - Fundação para a Ciência e a Tecnologia, within project LA/P/0063/2020. DOI10.54499/LA/P/0063/2020, https://doi.org/10.54499/LA/P/0063/2020.